# Efficient improvement of frequency-domain Kalman filter

Wenzhi Fan, Kai Chen, Jing Lu and Jiancheng Tao

*Abstract*—The frequency-domain Kalman filter (FKF) has been utilized in many audio signal processing applications due to its fast convergence speed and robustness. However, the performance of the FKF in under-modeling situations has not been investigated. This paper presents an analysis of the steady-state behavior of the commonly used diagonalized FKF and reveals that it suffers from a biased solution in under-modeling scenarios. Two efficient improvements of the FKF are proposed, both having the benefits of the guaranteed optimal steady-state behavior at the cost of a very limited increase of the computational burden. The convergence behavior of the proposed algorithms is also compared analytically. Computer simulations are conducted to validate the improved performance of the proposed methods.

## I. Introduction

THE Kalman filter has been widely used in many practical applications, such as spacecraft navigation, robot control and econometrics [1][2]. The frequency-domain Kalman filter (FKF) for acoustic echo cancelation (AEC) was developed in [3], utilizing a stochastic state-space model of the acoustic echo path formulated in the frequency-domain entirely. The FKF was further developed [4]-[7] and its application has been extended to dereverberation and acoustic feedback cancellation [8]-[9].

Compared with the normally used frequency-domain adaptive filters (FDAF) [10], the FKF does not require additional regularization or control mechanisms and is computationally efficient and inherently robust [3]. It is generally assumed that the adaptive filter is of sufficient filter length [3]-[7]. However, in many practical applications, the impulse response of the system can be extremely long [11]-[14], resulting in under-modeling situations. Therefore, it is meaningful to investigate the performance of the FKF when the filter is of deficient length.

It has been noticed that the Kalman filter provides a unifying framework for different types of adaptive transversal filters [15] and it is indicated in [16] that the optimal solution of the Kalman filter is the same as the Wiener solution with white observation noise, finite-dimensional signal model and stationary process. In this paper, the steady-state behavior of the FKF is analyzed by investigating the optimal solution of the equivalent weight vector in time-domain. It is found that the FKF converges to a biased steady-state solution when the filter is of deficient length and the performance might deteriorate considerably. The FKF can be understood as a variable step-size FDAF [3] which also suffers from similar problems [13]-[19]. To resolve performance deterioration, two efficient improvements of the FKF are proposed, leading to guaranteed optimal steady-state behavior. Convergence behavior of both methods are compared, and simulations are carried out to verify their performance. Throughout this paper, lowercase letters are used for time-domain signals, uppercase letters mostly for frequency-domain signals with a few annotated exceptions, and bold letters for vectors or matrices.

## II. Analysis of The Steady-state Behavior

The basic structure of the FKF [3] is briefly revisited. Let $\mathbf{w}(k)=[w_0(k), …, w_{N-1}(k)]^T$ be the filter coefficients of length $N$, where $k$ denotes the frame index, and the superscript T denotes the transpose operation. Similarly, let $\mathbf{d}(k)=[d(kN–N+1), …, d(kN)]^T$ be the desired signal vector, $\mathbf{s}(k)=[s(kN–N+1), …, s(kN)]^T$ be the observation noise vector (the near-end signal in AEC), and $\mathbf{x}(k)=[x(kN–M+1), …, x(kN)]^T$ be the reference signal vector, where $M$ denotes the frame size and $M=2N$.

The reference signal matrix in the frequency-domain can be denoted as $\mathbf{X}(k)=\mathrm{diag}\{\mathbf{Fx}(k)\}$, where $\mathbf{F}$ represents the Fourier transform matrix of size $M \times M$ and diag{ ·} creates a diagonal matrix from its input. Let $\mathbf{W}(k)=[W_0(k), …, W_{M-1}(k)]^T=\mathbf{F}[\mathbf{w}^T(k), \mathbf{0}_{1\times N}]^T$ be the frequency-domain filter coefficients, where $\mathbf{0}_{1\times N}$ is an all-zero vector of size $1\times N$, then the desired signal vector can be expressed in the frequency-domain as

$$\mathbf{D}(k) = \mathbf{G}_{0,N}\mathbf{X}(k)\mathbf{W}(k) + \mathbf{S}(k), \quad (1)$$

where $\mathbf{D}(k)=\mathbf{F}[\mathbf{0}_{1\times N}, \mathbf{d}^T(k)]^T$ is the frequency-domain desired signal vector, $\mathbf{S}(k)=\mathbf{F}[\mathbf{0}_{1\times N}, \mathbf{s}^T(k)]^T$ is the frequency-domain observation noise, and

$$\mathbf{G}_{0,N} = \mathbf{F}\begin{bmatrix}\mathbf{0}_{N\times N} & \mathbf{0}_{N\times N} \\ \mathbf{0}_{N\times N} & \mathbf{I}_{N\times N}\end{bmatrix}\mathbf{F}^{-1}. \quad (2)$$

A first order statistical Markov model is used to describe the time-varying property of the unknown system [15]:

$$\mathbf{W}(k+1) = A\cdot\mathbf{W}(k) + \Delta\mathbf{W}(k), \quad (3)$$

where $A$ is the transition parameter in the range $0<A\leq 1$ and $\Delta\mathbf{W}(k)$ is the process noise vector with covariance matrix

This work was supported by the National Natural Science Foundation of China (Grant No. 11874219 and 11874218).

W. Fan, K. Chen, J. Lu and J. Tao are with the Key Laboratory of Modern Acoustics and Institute of Acoustics, Nanjing University, Nanjing 210093, China (e-mail: wenzhi_fan@ smail.nju.edu.cn; chenkai@nju.edu.cn; lujing@nju.edu.cn; jctao@nju.edu.cn).

$\Psi_{\Delta\Delta}(k) = \mathrm{diag}\{[\Psi_{\Delta\Delta,0}(k), \ldots, \Psi_{\Delta\Delta,M-1}(k)]^T\}$. The state-space model for Kalman filter is formed by (1) and (3), which are respectively the observation equation and the state equation.

In order to decrease the computational complexity, the diagonalized version of the FKF was proposed in [3] as:

$$\mathbf{W}(k+1) = A\left[\mathbf{W}(k) + \mathbf{G}_{N,0}\mathbf{K}(k)\mathbf{E}(k)\right], \quad (4)$$

$$\mathbf{K}(k+1) = \mathbf{P}(k)\mathbf{X}^H(k) \cdot \left[\mathbf{X}(k)\mathbf{P}(k)\mathbf{X}^H(k) + M \cdot \mathrm{diag}\{\Phi_{SS}(k)\}\right]^{-1}, \quad (5)$$

$$\mathbf{P}(k+1) = A^2\left[\mathbf{I}_{M\times M} - (N/M)\mathbf{K}(k)\mathbf{X}(k)\right]\mathbf{P}(k) + M \cdot \mathrm{diag}\{\Phi_{\Delta\Delta}(k)\}, \quad (6)$$

where $\mathbf{E}(k)$ is the frequency-domain error vector, $\mathbf{K}(k)$ is the Kalman gain, the superscript H represents the conjugate transpose operation, $\Phi_{\Delta\Delta}(k)$ and $\Phi_{SS}(k)$ are the power spectral density of the process noise and observation noise respectively, $\mathbf{P}(k) = \mathrm{diag}\{[P_0(k), \ldots, P_{M-1}(k)]^T\}$ is the state estimation error covariance matrix based on Kalman filter theory [15][16], and $\mathbf{G}_{N,0}$ is the constraining matrix with the form:

$$\mathbf{G}_{N,0} = \mathbf{F}\begin{bmatrix} \mathbf{I}_{N\times N} & \mathbf{0}_{N\times N} \\ \mathbf{0}_{N\times N} & \mathbf{0}_{N\times N} \end{bmatrix}\mathbf{F}^{-1}. \quad (7)$$

It is also demonstrated in [3] that there is a fundamental relationship between the diagonalized FKF and the FDAF. Consequently, the updating equation of the FKF can be written in a FDAF form with simple substitution:

$$\mathbf{W}(k+1) = A\left[\mathbf{W}(k) + \mathbf{G}_{N,0}\mathrm{diag}\{\mu(k)\}\mathbf{X}^H(k)\mathbf{E}(k)\right], \quad (8)$$

where the step-size matrix of the FKF can be described as

$$\mathrm{diag}\{\mu(k)\} = \mathbf{P}(k)\left[\mathbf{X}(k)\mathbf{P}(k)\mathbf{X}^H(k) + M \cdot \mathrm{diag}\{\Phi_{SS}(k)\}\right]^{-1}. \quad (9)$$

To analyze the FKF, multiplying both sides of (8) by $\mathbf{F}^{-1}$ yields

$$\begin{bmatrix} \mathbf{w}(k+1) \\ \mathbf{0}_{N\times 1} \end{bmatrix} = A\left[\begin{bmatrix} \mathbf{w}(k) \\ \mathbf{0}_{N\times 1} \end{bmatrix} + \begin{bmatrix} \mathbf{I}_{N\times N} & \mathbf{0}_{N\times N} \\ \mathbf{0}_{N\times N} & \mathbf{0}_{N\times N} \end{bmatrix}\mathbf{M}(k)\mathbf{X}_C(k)\begin{bmatrix} \mathbf{0}_{N\times 1} \\ \mathbf{e}(k) \end{bmatrix}\right], \quad (10)$$

where $\mathbf{e}(k) = [e(kN-N+1), \ldots, e(kN)]^T$,

$$\mathbf{X}_C(k) = \mathbf{F}^{-1}\mathbf{X}(k)\mathbf{F} = \begin{bmatrix} \mathbf{X}_{C,1}(k) & \mathbf{X}_{C,2}(k) \\ \mathbf{X}_{C,2}(k) & \mathbf{X}_{C,1}(k) \end{bmatrix} \quad (11)$$

is a circulant matrix whose first row is $\mathbf{x}(k)$ and

$$\mathbf{M}(k) = \mathbf{F}^{-1}\mathrm{diag}\{\mu(k)\}\mathbf{F} = \begin{bmatrix} \mathbf{M}_1(k) & \mathbf{M}_2(k) \\ \mathbf{M}_2(k) & \mathbf{M}_1(k) \end{bmatrix} \quad (12)$$

is also a circulant matrix whose first row is $\mathbf{F}^{-1}\mu(k)$. $\mathbf{X}_{C,1}(k)$, $\mathbf{X}_{C,2}(k)$, $\mathbf{M}_1(k)$ and $\mathbf{M}_2(k)$ are matrices with size $N\times N$.

Substitute (11) and (12) into (10), the time-domain update equation for the FKF can be described as

$$\mathbf{w}(k+1) = A\mathbf{w}(k) + A\left[\mathbf{M}_1(k)\mathbf{X}_{C,2}(k) + \mathbf{M}_2(k)\mathbf{X}_{C,1}(k)\right]\mathbf{e}(k), \quad (13)$$

where

$$\mathbf{e}(k) = \mathbf{d}(k) - \mathbf{X}_{C,2}^T(k)\mathbf{w}(k). \quad (14)$$

To analyze the convergence behavior of the system, the reference signal and the filter coefficients are regarded independent, which is a common assumption in adaptive filter analysis [11]. Furthermore, the step-size vector $\mu(k)$, as well as its related matrix $\mathbf{M}(k)$, is assumed to be independent of the reference signal and the filter coefficients, since $\mu(k)$ varies slowly as the algorithm approaches the steady state [7]. Such assumption is widely adopted in the analysis of variable step-size adaptive algorithm [20]-[25]. The mean convergence behavior of the time-domain filter coefficients can be determined by taking expectation on both sides of (13) as

$$\mathrm{E}\{\mathbf{w}(k+1)\} = A\left[\mathbf{I}_{N\times N} - \Lambda_1(k)\mathbf{R} - \Lambda_2(k)\hat{\mathbf{R}}\right]\mathrm{E}\{\mathbf{w}(k)\} + A\left[\Lambda_1(k)\mathbf{r} + \Lambda_2(k)\hat{\mathbf{r}}\right], \quad (15)$$

with

$$\mathbf{R} = \mathrm{E}\{\mathbf{X}_{C,2}(k)\mathbf{X}_{C,2}^T(k)\} = N\mathbf{R}_x, \hat{\mathbf{R}} = \mathrm{E}\{\mathbf{X}_{C,1}(k)\mathbf{X}_{C,2}^T(k)\},$$
$$\mathbf{r} = \mathrm{E}\{\mathbf{X}_{C,2}(k)\mathbf{d}(k)\} = N\mathbf{r}_{xd}, \hat{\mathbf{r}} = \mathrm{E}\{\mathbf{X}_{C,1}(k)\mathbf{d}(k)\}, \quad (16)$$
$$\Lambda_1(k) = \mathrm{E}\{\mathbf{M}_1(k)\}, \Lambda_2(k) = \mathrm{E}\{\mathbf{M}_2(k)\},$$

where $\mathbf{R}_x$ represents the autocorrelation matrix of the reference signal and $\mathbf{r}_{xd}$ represents the correlation vector between the reference signal and the desired signal. The steady-state solution of (15) can be obtained as

$$\mathrm{E}\{\mathbf{w}(\infty)\} = A\left[(1-A)\mathbf{I}_{N\times N} + A\Lambda_1(\infty)\mathbf{R} + A\Lambda_2(\infty)\hat{\mathbf{R}}\right]^{-1} \cdot \left[\Lambda_1(\infty)\mathbf{r} + \Lambda_2(\infty)\hat{\mathbf{r}}\right]. \quad (17)$$

For the situation of a sufficient filter length, i.e. the length of the unknown system $L \leq N$, the desired signal vector can be described as

$$\mathbf{d}(k) = \mathbf{X}_{C,2}^T(k)\left[w_0, \ldots, w_{L-1}, \mathbf{0}_{1\times(N-L)}\right]^T + \mathbf{s}(k). \quad (18)$$

It can be easily verified that

$$\mathbf{r} = \mathbf{R}\left[w_0, \ldots, w_{L-1}, \mathbf{0}_{1\times(N-L)}\right]^T,$$
$$\hat{\mathbf{r}} = \hat{\mathbf{R}}\left[w_0, \ldots, w_{L-1}, \mathbf{0}_{1\times(N-L)}\right]^T. \quad (19)$$

If the transition parameter $A$ is set to be 1, the steady-state solution can be written as $\mathrm{E}\{\mathbf{w}(\infty)\} = [w_0, \ldots, w_{L-1}, \mathbf{0}_{1\times(N-L)}]^T$ by substituting (19) into (17), which means that the FKF achieves a perfect match between the adaptive filter and the unknown system.

However, when the adaptive filter is of deficient length, the term in (13), $\mathbf{M}_2(k)\mathbf{X}_{C,1}(k)$, obstructs the filter convergence and $\mathrm{E}\{\mathbf{w}(\infty)\}$ in (17) cannot be simplified to the optimal solution, leading to performance deterioration of the FKF. Similar problems exist for the FBLMS algorithm, which has been addressed in [13]-[19].

### III. PROPOSED METHODS

#### A. The first modified FKF (MFKF1)

To circumvent the unfavorable effect of $\mathbf{M}_2(k)\mathbf{X}_{C,1}(k)$, the update equation can be revised by changing the position of the constraining matrix $\mathbf{G}_{N,0}$ as

$$\mathbf{W}(k+1) = A\left[\mathbf{W}(k) + \mathrm{diag}\{\mu(k)\}\mathbf{G}_{N,0}\mathbf{X}^H(k)\mathbf{E}(k)\right]. \quad (20)$$

Multiplying both sides of (20) by $\mathbf{F}^{-1}$ leads to

$$\begin{bmatrix} \mathbf{w}(k+1) \\ \mathbf{w}_{wr}(k+1) \end{bmatrix} = A\left[\begin{bmatrix} \mathbf{w}(k) \\ \mathbf{w}_{wr}(k) \end{bmatrix} + \mathbf{M}(k)\begin{bmatrix} \mathbf{I}_{N\times N} & \mathbf{0}_{N\times N} \\ \mathbf{0}_{N\times N} & \mathbf{0}_{N\times N} \end{bmatrix}\mathbf{X}_C(k)\begin{bmatrix} \mathbf{0}_{N\times 1} \\ \mathbf{e}(k) \end{bmatrix}\right], \quad (21)$$

where $\mathbf{w}_{wr}(k)$ represents the part of filter coefficients that suffers from the wraparound effect of circular convolution. Focusing on the causal part of the filter coefficients, the

following update equation in time-domain can be obtained as
$$\mathbf{w}(k+1) = A\mathbf{w}(k) + A\mathbf{M}_1(k)\mathbf{X}_{C,2}(k)\mathbf{e}(k). \quad (22)$$
Taking expectation on both sides of (22) yields
$$\mathrm{E}\{\mathbf{w}(k+1)\} = A[\mathbf{I}_{N\times N} - \mathbf{\Lambda}_1(k)\mathbf{R}]\mathrm{E}\{\mathbf{w}(k)\} + A\mathbf{\Lambda}_1(k)\mathbf{r}, \quad (23)$$
whose steady-state solution, $\mathrm{E}\{\mathbf{w}(\infty)\}=\mathbf{R}_x^{-1}\mathbf{r}_{xd}$, is the well-known optimal Wiener solution [11] when the transition parameter $A$ is set to be 1. The parameter $A$ depends on the variability of the unknown system and is usually a constant close to 1 in practical applications [3], where the steady-state behavior resembles the case when $A$ is set to be 1.

Comparing (8) with (20), the computational complexity of FKF and MFKF1 seems the same. However, extra constraints are needed to eliminate the influence of $\mathbf{w}_{wr}(k)$ in the proposed algorithm when computing the output, resulting in a limited increase of the computational load with merely one extra pair of FFT/IFFT.

*B. The second modified FKF (MFKF2)*

Intuitively the FKF converges to the optimal solution if the transition parameter $A$ is set to be 1 and the matrix $\mathbf{M}_2$ is an all-zero matrix. Therefore, another possible modification aims to transform the step-size matrix of the FKF into
$$\mathrm{diag}\{\boldsymbol{\mu}(k)\} = \xi(k)\mathbf{I}_{M\times M}, \quad (24)$$
with
$$\xi(k) = \min\left\{\mathbf{P}(k)\left[\mathbf{X}(k)\mathbf{P}(k)\mathbf{X}^H(k) + M\cdot\mathrm{diag}\{\Phi_{SS}(k)\}\right]^{-1}\right\}, \quad (25)$$
where min{·} picks the minimum value from the diagonal elements of the input matrix. In this case, (12) is simplified as
$$\mathbf{M}(k) = \mathbf{F}\mathrm{diag}\{\boldsymbol{\mu}(k)\}\mathbf{F}^{-1} = \xi(k)\mathbf{I}_{M\times M}. \quad (26)$$
Thus, the submatrix $\mathbf{M}_2$ is an all-zero matrix, resulting in time-domain update equation as
$$\mathbf{w}(k+1) = A\left[\mathbf{w}(k) + \xi(k)\mathbf{X}_{C,2}(k)\mathbf{e}(k)\right]. \quad (27)$$
By taking expectation on both sides of (27), it can be easily seen that
$$\mathrm{E}\{\mathbf{w}(k+1)\} = A\left[\mathbf{I}_{N\times N} - E\{\xi(k)\}\mathbf{R}\right]\mathrm{E}\{\mathbf{w}(k)\} + AE\{\xi(k)\}\mathbf{r}, \quad (28)$$
whose steady-state solution is also $\mathrm{E}\{\mathbf{w}(\infty)\}=\mathbf{R}_x^{-1}\mathbf{r}_{xd}$ with $A=1$.

This modification is much simpler than the first approach, since it just requires an additional minimizing operation. Nevertheless, the lower computational complexity is at a cost of potentially slower convergence speed, which will be investigated subsequently.

*C. Analysis and comparisons of the proposed methods*

To analyze the convergence behavior of the MFKF1, it is assumed that the unknown system is of length $N$ and the transition parameter $A$ is 1. Define the frequency-domain filter coefficient error vector to be
$$\mathbf{V}(k) = \mathbf{W}(k) - \mathbf{W}_o, \quad (29)$$
where $\mathbf{W}_o$ is the optimal solution. Substituting (29) into (20) yields
$$\mathbf{V}(k+1) = \left[\mathbf{I}_{M\times M} - \mathrm{diag}\{\boldsymbol{\mu}(k)\}\mathbf{G}_{N,0}\mathbf{X}^H(k)\mathbf{G}_{0,N}\mathbf{X}(k)\right] \\ \cdot \mathbf{V}(k) + \mathrm{diag}\{\boldsymbol{\mu}(k)\}\mathbf{G}_{N,0}\mathbf{X}^H(k)\mathbf{E}_o(k), \quad (30)$$
where $\mathbf{E}_o(k)$ is the minimum error vector when $\mathbf{W}(k)$ is replaced by $\mathbf{W}_o$. Taking expectation on both sides of (30) leads to
$$\mathrm{E}\{\mathbf{V}(k+1)\} = \left[\mathbf{I}_{M\times M} - E\{\mathrm{diag}\{\boldsymbol{\mu}(k)\}\}\mathbf{G}_{N,0}\mathbf{R}_{XF}\right]\cdot\mathrm{E}\{\mathbf{V}(k)\}, \quad (31)$$
with
$$\mathbf{R}_{XF} = \mathrm{E}\{\mathbf{X}^H(k)\mathbf{G}_{0,N}\mathbf{X}(k)\}. \quad (32)$$
Note that the independence between $\mathbf{V}(k)$ and $\mathbf{X}(k)$ and the orthogonality between $\mathbf{X}$ and $\mathbf{E}_o$ [11] are both assumed here. And the step-size vector $\boldsymbol{\mu}(k)$ is assumed to be independent of $\mathbf{V}(k)$ and $\mathbf{X}(k)$ [20]-[25].

It has been shown in [26] that when the filter length $N$ and the frame size $M$ are sufficiently large, $\mathbf{R}_{XF}$ can be approximated as
$$\mathbf{R}_{XF} \approx N\cdot\mathrm{diag}\left\{\Phi_{xx,0}(e^{j\frac{2\pi\times 0}{M}}),...,\Phi_{xx,M-1}(e^{j\frac{2\pi\times(M-1)}{M}})\right\}, \quad (33)$$
where $\Phi_{xx}$ is the power spectral density of the reference signal. Furthermore, the constraining matrix can be approximated as [27]
$$\mathbf{G}_{N,0} \approx (N/M)\mathbf{I}_{M\times M}. \quad (34)$$
Recalling that all the matrices in (9) are diagonal, it can be easily found that the step-size of the FKF for each frequency bin is
$$\mu_i(k) = 1/\left[|X_i(k)|^2 + M\cdot\Phi_{SS,i}(k)/P_i(k)\right]. \quad (35)$$
As mentioned in Sec. I, $P_i(k)$ is the state estimation error covariance [3][16], therefore it is reasonable to assume that the value of $P_i(k)$ is large at the early stage of convergence. Hence, the second term of (35) is relatively insignificant compared to $|X_i(k)|^2$, leading to the following approximation:
$$\mu_i(k) \approx 1/|X_i(k)|^2 = 1/\left[M\cdot\Phi_{xx,i}(e^{j2\pi i/M})\right]. \quad (36)$$
Combining (33), (34) and (36), (31) can be simplified as
$$\mathrm{E}\{\mathbf{V}(k+1)\} \approx \left[1-(N/M)^2\right]\cdot\mathrm{E}\{\mathbf{V}(k)\}, \quad (37)$$
which indicates an exponentially fast convergence speed at the early stage of convergence.

For the MFKF2, the following approximation can be obtained by substituting (36) into (25):
$$\xi(k) = \min_i\{\mu_i(k)\} \approx 1/\max_i\{M\cdot\Phi_{xx,i}(e^{j2\pi i/M})\}. \quad (38)$$
Likewise, substituting (24), (29), (33), (34) and (38) into (8) and taking expectation on both sides of the equation yields
$$\mathrm{E}\{\mathbf{V}(k+1)\} \approx \left[\mathbf{I}_{M\times M} - \frac{N^2}{M^2}\frac{1}{\max_i\{\Phi_{xx,i}(e^{j2\pi i/M})\}}\right. \\ \left.\cdot\mathrm{diag}\left\{\Phi_{xx,0}(e^{j\frac{2\pi\times 0}{M}}),...,\Phi_{xx,M-1}(e^{j\frac{2\pi\times(M-1)}{M}})\right\}\right]\cdot\mathrm{E}\{\mathbf{V}(k)\}. \quad (39)$$
Since it is obvious that
$$\Phi_{xx,i}(e^{j2\pi i/M})/\max_i\{\Phi_{xx,i}(e^{j2\pi i/M})\} \leq 1, \quad (40)$$
it can be concluded that the convergence speed of the second proposed algorithm is slower than the first one at early stage of convergence by comparing (37) with (39).

## IV. COMPUTER SIMULATIONS

Computer simulations are carried out to verify the theoretical results and demonstrate the efficacy of the proposed algorithms.

signal through a 16-tap FIR filter. The length of the adaptive filter $N$ is 10 and the frame-length $M$ is 20 accordingly. The transition parameter $A$ is set to be 1.

Fig. 1(a) depicts the misalignments of the FKF, MFKF1 and MFKF2 in this under-modelling situation. The FKF converges with the fastest speed but to a biased steady-state solution, whereas the misalignments of both MFKF1 and MFKF2 are significantly smaller than the FKF. It is noted that the misalignment curve of the FKF is much smoother than that of the proposed algorithms, since its fluctuation is masked by the comparatively large deviation of the steady-state solution in the logarithmic coordinate. As analyzed in section III, it can be clearly seen that the convergence speed of the MFKF1 is faster than the MFKF2 in such circumstances, since the MFKF2 selects the smallest step-size conservatively. The steady-state filter coefficients are shown in Fig. 1(b). While the steady-state solution of the FKF differs from the Wiener solution, the proposed algorithms converge to the optimal solution perfectly.

Fig. 2 shows the misalignment curves of the MFKF1 with different values of $A$ in the under-modeling situation whose setup is the same as the above example. It has been pointed out that the parameter $A$ has influence on the convergence rate, the tracking ability and the steady-state misalignment [3][7]. It can be seen from Fig. 2 that the steady-state misalignment of the MFKF1 increases as the parameter $A$ decreases, but overall the MFKF1 with different values of $A$ performs significantly better than the standard FKF in this situation.

### B. A practical AEC example

Fig. 3 depicts the misalignment curves of the FKF and the MFKF in the actual AEC scenario. The echo signal is simulated by convolving the reference signal (clean speech) with a measured room impulse response in an office with a reverberation time of about 600 ms. The sampling rate is 16 kHz. The length of the adaptive filter $N$ is 512, which is significantly deficient for modeling the impulse response. The transition parameter $A$ is also set to be 1. It can be seen from Fig. 3 that the FKF converges faster at the initial stage, but the steady-state solution is obviously biased. The MFKF1 and MFKF2 both have a significantly better echo attenuation level, while the convergence speed of the MFKF1 is faster than the MFKF2, indicating a preferable performance in practical applications.

## V. CONCLUSIONS

The steady-state behavior of the diagonalized frequency-domain Kalman filter has been investigated in this paper. It is found that the steady-state solution of the FKF is not optimal in the under-modeling situation. On the basis of the analysis, two methods are proposed to improve the steady-state performance of the FKF. Both methods can guarantee an optimal steady-state solution with limited extra computational load, while the MFKF1 has a comparatively faster convergence speed than the MFKF2. Simulations on a simple system identification and a practical AEC system validate the efficacy of the proposed algorithms.

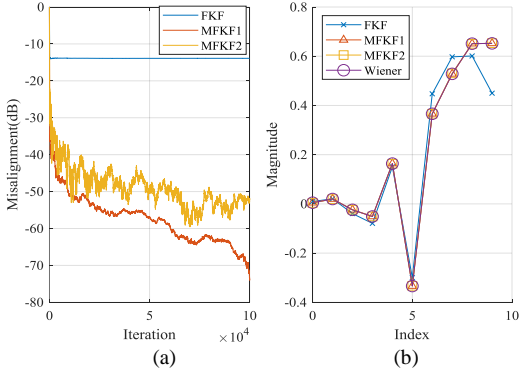

Fig. 1. (a) Misalignments of the FKF, MFKF1 and MFKF2. (b) Steady-state solution of the filter coefficients in the under-modeling example.

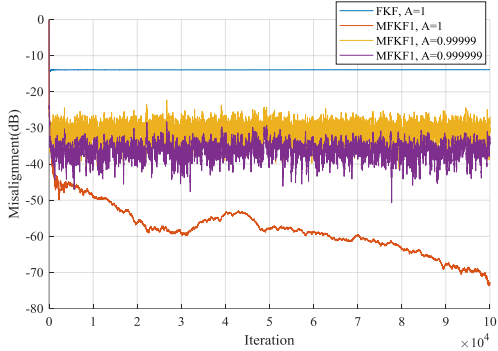

Fig. 2. Misalignments of FKF and MFKF1 with different transition parameters in the under-modeling example.

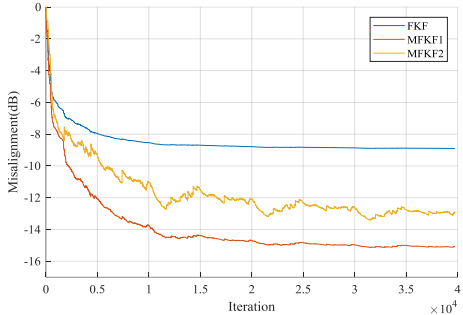

Fig. 3. Misalignments of the FKF, MFKF1 and MFKF2 in the actual AEC scenario.

Firstly, the performance of the analyzed algorithms is verified with a simple system identification example, whose setups are the same as the simulation for under-modeling situations in [14]. Then the convergence behavior of the proposed algorithm with different transition parameters is compared. Eventually, a practical AEC situation is considered and simulated. To demonstrate the convergence of algorithms, the normalized misalignment of the filter coefficients (in dB) is defined as

$$m(k) = 10\log_{10}\left[(\mathbf{w}(k)-\mathbf{w}_o)^T(\mathbf{w}(k)-\mathbf{w}_o)/\|\mathbf{w}_o\|^2\right], \quad (41)$$

with $\mathbf{w}_o$ is the optimal solution in time-domain, which is similar to the definition in [7].

### A. An illustrating example

In this case, the reference signal is generated by passing Gaussian white noise with unit variance through a 4-tap FIR filter. The desired signal is generated by passing the reference